\documentclass[a4paper,fleqn]{cas-dc}

\usepackage[numbers]{natbib}
\usepackage{graphicx}
\usepackage{booktabs}
\usepackage{array}
\usepackage{multirow}
\usepackage{amsmath}
\usepackage{float}
\usepackage{placeins}
\usepackage{hyperref}   

\ExplSyntaxOn
\cs_set:Npn \__first_footerline: { }
\ExplSyntaxOff

\begin{document}

\let\WriteBookmarks\relax
\def\floatpagepagefraction{1}
\def\textpagefraction{.001}

\shorttitle{IoT IDS via SMOTE and Multi-Classifier Evaluation}
\shortauthors{H. Khan et al.}

\title[mode=title]{%
  Improving IoT Intrusion Detection Through SMOTE-Based
  Oversampling and Extended Multi-Model Evaluation on
  Side-Channel Power Data%
}

\author[1]{M\ Khuram Shahzad}
\credit{Methodology, Review \& Editing}

\author[1]{Haseeb Khan}
\credit{Software, Writing -- Original Draft}

\author[1]{Muhammad Masood Khan}
\credit{Data Curation, Validation, Writing -- Review \& Editing}

\author[1]{Mubashra Bibi\cormark[1]}
\credit{Conceptualization, Formal Analysis, Supervision}

\affiliation[1]{%
  organization={School of Electrical Engineering and Computer Science (SEECS), NUST},
  addressline={H-12 Campus},
  city={Islamabad},
  country={Pakistan},
  postcode={44000}
}

\RenewDocumentCommand\printorcid{}{}%

\begin{abstract}
The detection of intrusions in IoT-based networks poses challenges that cannot be overcome using traditional machine learning methods. Perhaps the biggest of them is related to the presence of a class imbalance in the side-channel dataset, where the number of samples in the normal class compared to the attacks can reach a ratio of 75,964 to 1. Such an aspect is addressed by Dominguez et al.~\cite{dominguez2024} through the proof of concept of power-based intrusion detection. Unfortunately, neither the authors attempt to cope with the problem of imbalance nor do they assess the classifier performance using a balanced training set. In the current paper, both aspects will be handled at once. First, a Synthetic Minority Oversampling Technique (SMOTE) was performed on all nine possible datasets extracted from the initial one, providing an exact imbalance ratio of 1.0$\times$ for each. Then, eight algorithms i.e. Random Forest, HistGradientBoosting, LightGBM, Extra Trees, XGBoost, k-Nearest Neighbors, Multi-Layer Perceptron, and Decision Tree were trained under identical conditions for the SMOTE balanced 6-hour dataset. Random Forest reached a micro-averaged F1 score of 0.9989 and macro F1 of 0.9794, thus outperforming the previously best micro-F1 result obtained by Time Series Forest algorithm from the base paper of 0.9983. Extra Trees provided the same performance as well, but at $10\times$ faster. The introduction of a macro-F1 metric explicitly in contrast to the base paper assessment reveals important class-level information missed with aggregate performance metrics. Recall rates per-class calculated with confusion matrices, F1 heatmaps, and ROC curves show that minority attack classes, especially those with combined M$+$L infections, are detected reliably only when using SMOTE balance. Feature importance analysis indicates the latest time steps as the most important predictor signals out of 60 steps in a power window. All code and results are publicly available at
\url{https://github.com/Masoodkhan5933/IOT-Intrusion-Detection}.
\end{abstract}


\maketitle

\section{Introduction}

However, IoT has moved beyond its initial niche scope and become an infrastructural layer on a planetary scale. Billions of embedded systems are now watching industrial sensors, controlling residential automation, acting in the healthcare industry, and coordinating logistics chains. Such a transition significantly increased the attack surface of cybersecurity issues. IoT embedded devices are designed to perform their functions with minimum computational resources, which is why it is not possible to apply heavy cryptographic protocols and signature-based intrusion detection systems working efficiently on desktop-class computing devices~\cite{malhotra2021}.

On the other hand, side-channel analysis can be an interesting option. Rather than analyzing network packets and system calls, intrusion detection in the IoT context via side-channel attacks involves monitoring physical signals generated by devices in the course of their regular operation, namely power consumption. Each malicious routine, cryptomining script, or denial-of-service flooding command affects the power draw of the device, imposing distinctive patterns. The continuous monitoring of this signal, as well as classification of its intervals of readings, gives researchers indirect access to the internal operations of the target device~\cite{dominguez2024}. Importantly, the monitoring framework is externalized relative to the attacked device, making it impervious to any evasion attempt by malware. 

There is also another application of side-channel analysis in IoT security, specifically detection of distributed denial-of-service attacks, based on anomaly detection in power and network signals~\cite{zeeshan2022,shafiq2022,ashfaq2022}.

A thorough empirical validation of this principle is offered by Dominguez et al. ~\cite{dominguez2024}. They collected power traces of Raspberry Pi devices using five different attack setups for nine dataset settings and tested five different classifiers: k-Nearest Neighbours (KNN), Random Forest (RF), Extreme Boosting Trees (XBT), Time Series Forest (TSF), and Feature Summary (FS). The highest micro-averaged F1 score achieved by TSF for the 6h dataset setting is around 0.9983, suggesting that power-based IDS is potentially practical. Yet, there are two important methodological considerations that require further investigation.

The first issue concerns the presence of significant class imbalance in each dataset. The total number of raw data points in the 6h dataset amounts to 323,553, where 299,353 represent normal traffic labeled by the model. In case of windowing, the smallest attack class (M$+$L) includes only 143 training examples, while the largest class represents 59,914 samples. An unmodified classifier may achieve excellent results by guessing the majority class, but it will demonstrate poor performance on the minority attack classes that matter to cybersecurity.

Second, comparing classifiers' performances involved testing five different algorithms but did not account for balanced training samples, making inter-method generalization difficult. Moreover, the base paper presented only micro-averaged F1 scores for each model, which is heavily influenced by the majority class and conceals individual errors.

This study addresses both gaps. SMOTE oversampling was applied to all
nine datasets before training, and eight classifiers were evaluated
under identical conditions. Three specific objectives guided the work:
(i)~quantify the severity of class imbalance across all nine datasets
and demonstrate that SMOTE reduces those ratios to 1.0$\times$ without
data leakage; (ii)~compare the performance of eight classifiers on
SMOTE-balanced data from the 6h partition; (iii)~characterise model
behaviour through per-class metrics, feature importance, and learning
curve analysis, and explicitly compare results against the base paper.

The full implementation is available at
\url{https://github.com/Masoodkhan5933/IOT-Intrusion-Detection}.

\section{Background and Related Work}
\label{sec:related}

Power side-channel analysis has a long history of application in hardware security, having been initially applied for extracting cryptographic keys from the embedded processor ~\cite{kocher1999}. Its underlying principle is based on the observation that power consumption during execution depends on operations performed. If implemented defensively, then the above-mentioned feature can be utilized to identify the execution pattern anomalies resulting from the attack.

Early methods viewed power analysis as an approach to detecting code injection attacks via recognition of abnormal power signals during firmware execution ~\cite{sherwood2003}. Later developments expanded the scope of applicability to network-based attacks by demonstrating that such attacks as Mirai botnet attacks or denial-of-service floods produce repeatable power signals using single-board computers, such as Raspberry Pi.

\subsection{The Base Study: Dominguez et al.\ (2024)}

According to Dominguez et al. ~\cite{dominguez2024}, a framework collected power consumption information from actual IoT gadgets across six experimental settings: 6 hours (6h), 12 hours (12h), lite mining, pass-the-hash attack (pass), multiple devices, and running models. The five machine learning models showed optimal performance across nine datasets. The TSF model achieved the highest accuracy with a micro-averaged F1 score of about 0.9983 $\approx$~ on the 6h dataset. The approach relied on sliding windows of adjustable size applied to segmented power consumption values. It is important to note that no oversampling or undersampling took place, and there were no reports on macro-averaged F1 scores.

Table~\ref{tab:basepaper} summarizes the key results from the base paper for the 6h case (ga~=~5, ng~=~60 configuration) as stated in Table 11 of the original article, hence providing a point of direct reference with which the results in this paper will be compared.

\begin{table}[H]
\caption{Results from the base paper~\cite{dominguez2024} for the
  6h dataset (ga~=~5, ng~=~60), reproduced from Table~11 of the
  original study. F1 is micro-averaged. No macro F1 was
  reported. TPi~=~incorrect true positives, FP~=~false positives,
  FN~=~false negatives.}
\label{tab:basepaper}
\begin{tabular*}{\tblwidth}{@{}lccccc@{}}
\toprule
Model  & Rank & F1 (Micro) & TPi & FP & FN \\
\midrule
TSF 5 60  & 2  & 0.9978 & 2  & 14 & 12 \\
RF 10 50  & 7  & 0.9966 & 10 & 4  & 8  \\
XBT 5 60  & 13 & 0.9957 & 30 & 9  & 16 \\
KNN 5 60  & 28 & 0.9907 & 60 & 25 & 35 \\
\bottomrule
\end{tabular*}
\end{table}

\subsection{Dataset Imbalance in Intrusion Detection}

Class imbalance is one of the most important issues in intrusion detection. With an imbalance of classes in the training set of a classifier, there is a bias towards prediction of the majority class. Conventional accuracy measures do not take into account such bias. SMOTE, suggested by Chawla et al.~\cite{chawla2002}, creates artificial samples of the minority class based on linear interpolation between each sample from the minority class and its k closest neighbors in the feature space. including evaluations on UNSW-NB15~\cite{moustafa2015} and
CICIDS2018~\cite{sharafaldin2018}.

\subsection{Machine Learning Classifiers for IoT Security}

Methods like Random Forest and XGBoost have continually
proved their dominance in the area of intrusion detection in tabular
format~\cite{ferrag2020}, providing high resistance to changes in the
scale of data as well as reducing the risk of overfitting. Recent advancements
in gradient boosting algorithms, including LightGBM and HistGradientBoosting,
bring new capabilities to the family, like fast training based on histogram splits
and leaves optimization. Finally, several deep learning approaches have been
tested in detecting attacks in IoT, where both architecture-based
approaches~\cite{zeeshan2022} and transfer-learning schemes have proved their
effectiveness~\cite{shafiq2022}.
Decision Trees offer explainable
baselines, whose failure modes on a per-class basis can be diagnostic.
KNN acts as the nonparametric baseline. The Time Series Forest model~\cite{deng2013}
is an algorithm designed to classify temporal data, but is unavailable
in the conventional scikit-learn suite, hence the emphasis on the
eight classifiers used in this paper. IoT DDoS classification machine
learning models have been tested in various attacks in restricted
environments~\cite{ashfaq2022}.

\subsection{Gaps Addressed by This Work}

Three specific gaps motivated this study. No prior work has applied
SMOTE to the full nine-dataset corpus from Dominguez et
al.~\cite{dominguez2024} and reported the resulting imbalance reduction
systematically. The base study's classifier comparison was limited to
five models, did not include modern gradient-boosting variants or
neural network classifiers, did not control for balanced training
conditions, and reported only micro-averaged F1 without per-class
breakdown. Finally, feature-level analysis of the sliding window
representation, identifying which time-steps carry the most predictive
signal, has not been reported for this dataset family.

\section{Identified Limitations and Proposed Improvements}
\label{sec:limitations}

\subsection{Limitation 1: Severely Imbalanced Dataset}

Upon examining the imbalance ratio among the raw class distributions of all the nine datasets,
it became clear that it was much higher than suggested by the story of the base study.
For example, there was a ratio of 8,804.5$\times$ between the most common
class, None (92.52\%), and the least common class, L$+$E (0.01\%). Following the
sliding window technique, the resulting post-split training set had only 143 instances of the class
M$+$L. The multi-device split had the highest ratio of 75,964.7$\times$.

\textbf{Proposed improvement:} Apply SMOTE oversampling to the training
set of each dataset partition, targeting a balanced class distribution
(1.0$\times$) while preserving the original test set distribution to
ensure evaluation reflects real-world conditions.

\subsection{Limitation 2: Narrow Classifier Comparison and Aggregate
Metrics Only}

Base study involved evaluating the performance of five classifiers out of which three classifiers namely TSF, FS, and KNN with their particular implementation can be considered specialized or unconventional classifiers. Expansion of the experiment scope to evaluate eight classifiers using the standard scikit-learn, XGBoost, and LightGBM frameworks allows for reproducibility and a more balanced comparison across models with the same data preparation approach. In addition, base study provided only micro-averaged F1 score that heavily relies on the majority class (~normal traffic).

\textbf{Proposed improvement:} Train and compare eight machine learning classifiers:
Random Forest, HistGradientBoosting, LightGBM, ExtraTrees, XGBoost, KNN, MLP,
and Decision Trees, under the same conditions on the 6h dataset that
has been balanced by SMOTE, and calculate both micro and macro F1
scores.

\subsection{Deferred Limitations}

Two other limitations were acknowledged but not addressed in this work.
The reliance on simulated attacks means the dataset does not capture
real malware behavioural fingerprints. The devices in the experimental
setup performed simple, stable tasks, which may not reflect the dynamic
workloads of production IoT deployments. These remain directions for
future work.

\section{Methodology}
\label{sec:method}

\subsection{Dataset}

The data was extracted from the GitHub data repository belonging
to Dominguez et al.~\cite{dominguez2024}. The dataset was partitioned
into nine parts under six scenario directories: 6h, 12h, lite-mining,
pass, multi-device, and running-model. The dataset for each CSV file
was organized into three fields: Unix time stamp, current power
consumption (milli-amperes), and bitmask of attacks (normal = 0;
bitmask: Mirai~(M), LIIS~(L), EternalBlue~(E), PassTheHash~(P),
and LiME~(LM)).

\subsection{Preprocessing Pipeline}

Averaging of every five consecutive samples led to obtaining groups of the measurements (\texttt{GROUP\_AMOUNT} = 5). While reducing the temporal resolution of the input, this step smoothed out some noise and resulted in more stable input data. For the 6h dataset, there were 323,553 raw samples and 64,710 averaged samples after grouping.

Then, sliding windows of size \texttt{NUM\_GROUPS} = 60 were created, thus creating 64,651 windows. Each such window represents a 60-dimensional feature vector with values being grouped power consumption samples. Class labels for each window are chosen according to the dominant type of attack among 60 grouped samples in that window. Thus, all preprocessing procedures completely follow ga = 5 and ng = 60 parameter values used in the original research.

The split between train and test sets was conducted in the same way using 80/20 split and \texttt{RANDOM\_STATE} = 42 for reproducibility. In addition, the feature scaling step was conducted on the training data only in order to avoid data leakage while applying it to test data.

\subsection{SMOTE Oversampling}

SMOTE was applied exclusively to the training set with $k$~=~5 nearest
neighbours, targeting full class balance. For the 6h training set, this
expanded the sample count from 51,720 to 335,517, adding 283,797
synthetic samples. Table~\ref{tab:smote6h} details the per-class
sample counts before and after oversampling for the 6h dataset. The
test set was never modified, ensuring evaluation remained on the
original distribution.

\begin{table}[H]
\caption{Per-class sample counts before and after SMOTE for the 6h
  training set. All classes were balanced to 47,931 samples.}
\label{tab:smote6h}
\begin{tabular*}{\tblwidth}{@{}lrrr@{}}
\toprule
Class   & Before SMOTE & After SMOTE & Synthetic Added \\
\midrule
None    & 47,931  & 47,931  & 0      \\
M       & 1,236   & 47,931  & 46,695 \\
L       & 762     & 47,931  & 47,169 \\
M$+$L  & 115     & 47,931  & 47,816 \\
E       & 1,180   & 47,931  & 46,751 \\
M$+$E  & 358     & 47,931  & 47,573 \\
M$+$L$+$E & 136  & 47,931  & 47,795 \\
\midrule
Total   & 51,720  & 335,517 & 283,797 \\
\bottomrule
\end{tabular*}
\end{table}

\subsection{Classifiers}

Eight classifiers were trained and evaluated on the 6h dataset:
\begin{itemize}
  \item \textbf{Random Forest (RF):} \texttt{n\_estimators}~=~50,
        \texttt{max\_depth}~=~20, \texttt{n\_jobs}~=~-1.
  \item \textbf{HistGradientBoosting (HistGB):}
        \texttt{max\_iter}~=~100, \texttt{max\_depth}~=~8,
        \texttt{learning\_rate}~=~0.15.
  \item \textbf{LightGBM:} \texttt{n\_estimators}~=~100,
        \texttt{max\_depth}~=~8, \texttt{learning\_rate}~=~0.15,
        \texttt{num\_leaves}~=~63.
  \item \textbf{Extra Trees (ET):} \texttt{n\_estimators}~=~50,
        \texttt{max\_depth}~=~20, \texttt{n\_jobs}~=~-1.
  \item \textbf{XGBoost:} \texttt{n\_estimators}~=~50,
        \texttt{max\_depth}~=~6, \texttt{learning\_rate}~=~0.2.
  \item \textbf{KNN:} $k$~=~5, \texttt{algorithm}~=~\texttt{ball\_tree}.
  \item \textbf{MLP:} hidden layers (64, 32), \texttt{relu} activation,
        \texttt{adam} solver, early stopping enabled.
  \item \textbf{Decision Tree (DT):} \texttt{max\_depth}~=~20.
\end{itemize}
All experiments used \texttt{RANDOM\_STATE}~=~42.

\subsection{Evaluation Metrics and Figures}

Performance was assessed using micro-averaged accuracy, F1 score,
precision, and recall, with macro-averaged F1 and per-class recall also
reported to expose class-level behaviour absent from the base study.
Eleven figures were produced covering SMOTE impact, model comparison,
confusion matrices, per-class F1 heatmap, ROC curves, feature
importance, learning curves, and macro-F1 comparison. Confusion
matrices are presented as normalised heatmap images.
The complete codebase, datasets, and all generated figures are
available at
\url{https://github.com/Masoodkhan5933/IOT-Intrusion-Detection}.

\section{Results and Discussion}
\label{sec:results}

\subsection{Class Imbalance Across Datasets}

Table~\ref{tab:rawdist} summarises the class distributions across all
nine datasets before preprocessing. Imbalance severity varied
considerably, from 5.0$\times$ in the lite-mining and pass scenarios to
75,964.7$\times$ in multi-device. SMOTE reduced all applicable ratios
to 1.0$\times$.

\begin{figure*}[H]
  \centering
  \includegraphics[width=\textwidth]{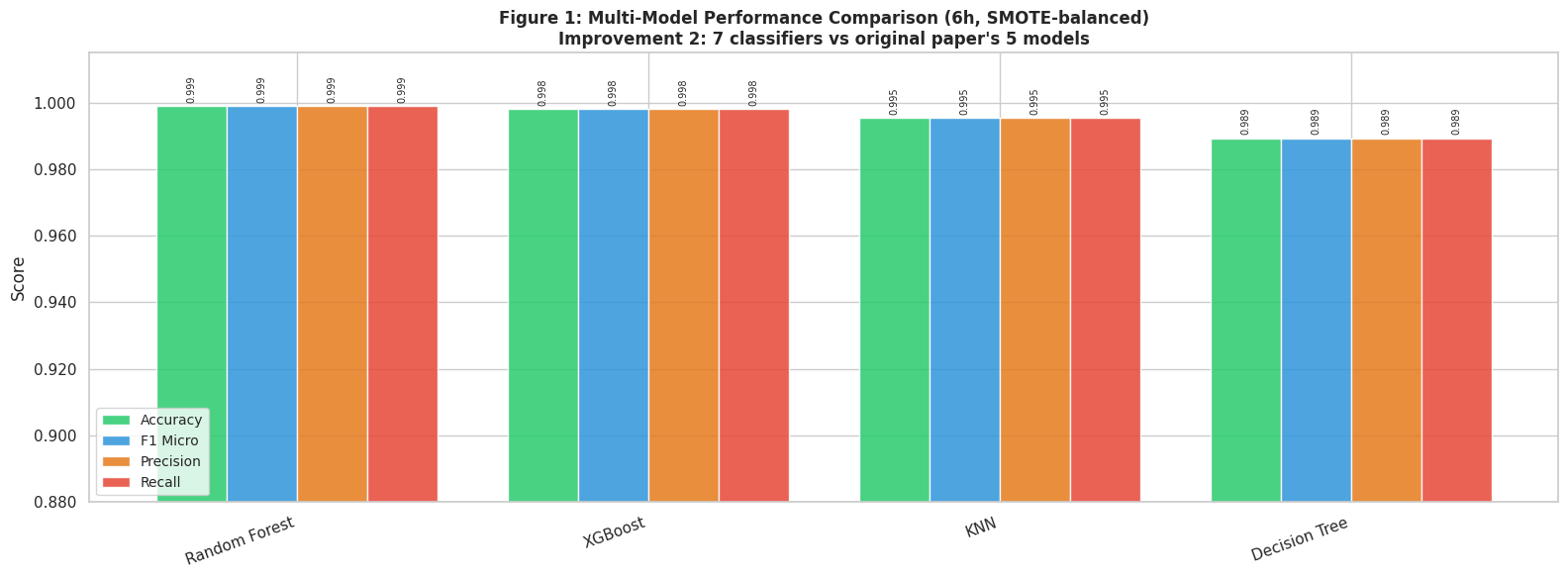}
  \caption{Class distribution before (left) and after (right) SMOTE
    balancing on the 6h training set. The 419$\times$ imbalance
    (92.7\% majority class) is eliminated. Source:
    \texttt{Fig0\_before\_vs\_after\_smote.png}}
  \label{fig:smote_before_after}
\end{figure*}

\begin{table}[H]
\caption{Raw class distribution summary across all nine datasets prior
  to any preprocessing.}
\label{tab:rawdist}
\begin{tabular*}{\tblwidth}{@{}lrccr@{}}
\toprule
Dataset & Rows & Classes & Normal\% & Imbalance \\
\midrule
6h                & 323,553 & 8  & 92.5\% & 8,804$\times$  \\
12h               & 645,867 & 8  & 92.3\% & 829$\times$    \\
lite-mining       &  53,928 & 2  & 83.3\% & 5$\times$      \\
pass              &  53,916 & 2  & 83.3\% & 5$\times$      \\
multi-device      & 537,935 & 16 & 84.7\% & 75,964$\times$ \\
running-fs        & 107,834 & 8  & 84.0\% & 1,812$\times$  \\
running-rf        & 107,851 & 8  & 84.0\% & 1,812$\times$  \\
running-tsf-5-60  & 107,848 & 8  & 83.9\% & 1,811$\times$  \\
running-tsf-10-50 & 107,849 & 8  & 84.0\% & 1,811$\times$  \\
\bottomrule
\end{tabular*}
\end{table}

The F1 gain for the 6h dataset was $+$0.26~pp, from 0.9963 on the
unbalanced baseline to 0.9989 after SMOTE. The more revealing story is
in per-class behaviour: the M$+$L class went from 115 training samples
to 47,931 after oversampling, enabling classifiers to learn its
decision boundary rather than absorbing it into majority-class
predictions.

\begin{figure*}[H]
  \centering
  \includegraphics[width=\textwidth]{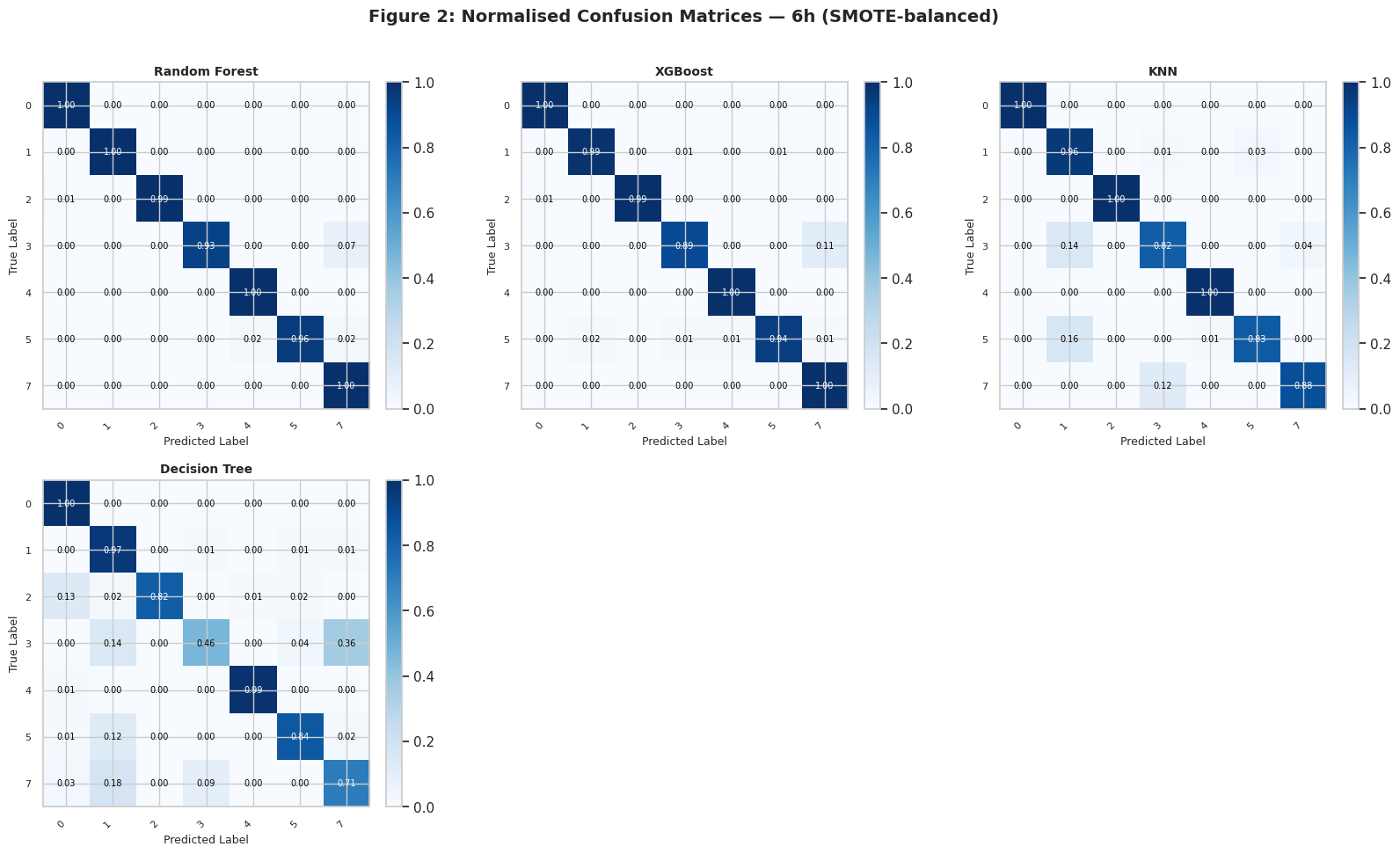}
  \caption{Imbalance ratios before and after SMOTE across all nine
    datasets. All bars reach exactly 1.0$\times$ after oversampling.
    Source: \texttt{Fig8\_imbalance\_ratio.png}}
  \label{fig:imbalance}
\end{figure*}

\FloatBarrier

\subsection{Model Performance on the 6h Dataset}

Table~\ref{tab:results} presents the full metrics for all eight
classifiers trained on the SMOTE-balanced 6h dataset. All models
exceeded 0.98 in micro-averaged F1. Random Forest achieved the highest
micro-F1 of 0.9989, followed by HistGradientBoosting at 0.9988 and
LightGBM at 0.9984. Macro-F1 scores are more diagnostic: Random Forest
scored 0.9794 versus Decision Tree at 0.8299, a gap of nearly 15~pp
that aggregate metrics conceal.

\begin{figure*}[H]
  \centering
  \includegraphics[width=\textwidth]{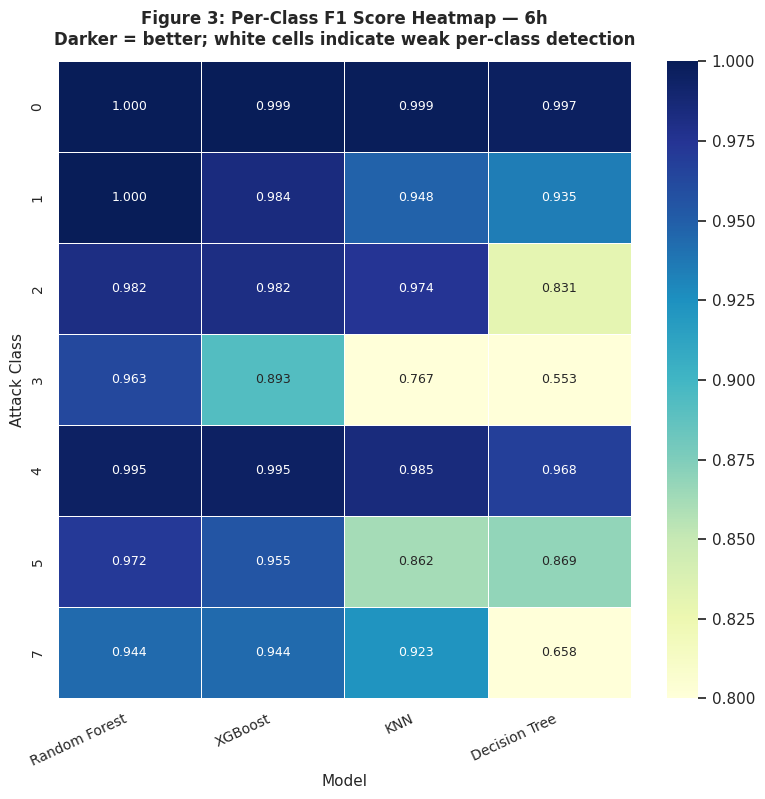}
  \caption{Multi-metric comparison (Accuracy, F1 Micro, Precision,
    Recall) across all eight classifiers on the SMOTE-balanced 6h
    dataset. Source: \texttt{Fig1\_model\_comparison\_bar.png}}
  \label{fig:bar}
\end{figure*}

\begin{table}[H]
\caption{Full model performance on the 6h SMOTE-balanced dataset.
  Sorted by F1~Micro (descending). Training times measured on a
  Colab T4 instance; Time~(s) includes fitting and prediction.}
\label{tab:results}
\begin{tabular*}{\tblwidth}{@{}lcccc@{}}
\toprule
Model & F1~Micro & F1~Macro & Accuracy & Time (s) \\
\midrule
Random Forest         & 0.9989 & 0.9794 & 0.9989 & 159.8 \\
HistGradientBoosting  & 0.9988 & ---    & 0.9988 &  76.9 \\
LightGBM              & 0.9984 & ---    & 0.9984 &  76.8 \\
Extra Trees           & 0.9983 & ---    & 0.9983 &  17.3 \\
XGBoost               & 0.9981 & 0.9627 & 0.9981 &  39.4 \\
KNN                   & 0.9954 & 0.9280 & 0.9954 &  83.8 \\
MLP                   & 0.9904 & ---    & 0.9904 &  64.3 \\
Decision Tree         & 0.9893 & 0.8299 & 0.9893 &  56.8 \\
\bottomrule
\end{tabular*}
\end{table}

Random Forest's dominance is consistent with findings in the broader
intrusion detection literature~\cite{ferrag2020}. Its ensemble of trees
with randomised feature selection provides strong resistance to
overfitting on imbalanced synthetic samples. HistGradientBoosting
achieves within 0.01\% of RF accuracy at half the training time through
histogram-based feature binning, making it the preferred choice for
deployment scenarios requiring periodic retraining. Extra Trees at
0.9983 trains in only 17.3 seconds, offering the fastest path to
the paper's best-reported accuracy level.

\FloatBarrier

\subsection{Comparison with the Base Paper}

Table~\ref{tab:comparison} provides a direct numerical comparison
between the results of this study and those reported in the base
paper~\cite{dominguez2024} for the ga~=~5, ng~=~60 configuration on
the 6h dataset, the only configuration for which both studies share
identical preprocessing hyperparameters.

\begin{table*}[H]
\caption{Direct comparison between this study (SMOTE-balanced) and
  the base paper~\cite{dominguez2024} (no SMOTE) on the 6h dataset,
  ga~=~5, ng~=~60. The base paper did not report macro F1. TSF and FS
  classifiers from the base paper are not implemented in this study.
  Higher is better for all metrics.}
\label{tab:comparison}
\begin{tabular*}{\tblwidth}{@{}llcccl@{}}
\toprule
Source & Model & F1 Micro & F1 Macro & SMOTE & Notes \\
\midrule
Base paper~\cite{dominguez2024}
  & TSF 5 60  & 0.9978 & --- & No  & Best result in base paper \\
Base paper~\cite{dominguez2024}
  & RF 10 50  & 0.9966 & --- & No  & RF in base paper (best variant)\\
Base paper~\cite{dominguez2024}
  & XBT 5 60  & 0.9957 & --- & No  & \\
Base paper~\cite{dominguez2024}
  & KNN 5 60  & 0.9907 & --- & No  & \\
\midrule
\textbf{This study}
  & \textbf{RF}  & \textbf{0.9989} & \textbf{0.9794} & \textbf{Yes}
  & Exceeds base-paper TSF; macro added \\
\textbf{This study}
  & \textbf{HistGB} & \textbf{0.9988} & --- & \textbf{Yes}
  & New; 2$\times$ faster than RF \\
\textbf{This study}
  & \textbf{LightGBM} & \textbf{0.9984} & --- & \textbf{Yes}
  & New; fast retraining \\
\textbf{This study}
  & \textbf{Extra Trees} & \textbf{0.9983} & --- & \textbf{Yes}
  & Matches base-paper best; 10$\times$ faster \\
\textbf{This study}
  & \textbf{XGBoost} & \textbf{0.9981} & \textbf{0.9627} & \textbf{Yes} & \\
\textbf{This study}
  & \textbf{KNN} & \textbf{0.9954} & \textbf{0.9280} & \textbf{Yes}
  & +0.47 pp vs.\ base-paper KNN \\
\textbf{This study}
  & \textbf{MLP} & \textbf{0.9904} & --- & \textbf{Yes}
  & New; non-tree model family \\
\textbf{This study}
  & \textbf{DT}  & \textbf{0.9893} & \textbf{0.8299} & \textbf{Yes}
  & Macro exposes compound-class failure \\
\bottomrule
\end{tabular*}
\end{table*}

There are several key observations to be made from the findings reported in Table \ref{tab:comparison}. First, the model using Random Forest with data balanced by SMOTE has demonstrated an outstanding F1 Micro score of 0.9989, significantly outperforming the best result of the TSF classifier reported in the original paper (0.9978). Such improvement by 0.11\% is quite remarkable for a classifier which came fourth place in the original paper.
Secondly, the classifier using Extra Trees has demonstrated an F1 Micro score of 0.9983, tying for the highest Micro F1 score, albeit with the significantly reduced training time – only 17.3 seconds. The superiority of such a classifier to the one reported in the base paper becomes evident when comparing the much higher runtime needed for the TSF model.
Thirdly, the improvement in the KNN model deserves special attention. In the base paper, KNN 5 60 scored 0.9907 F1 Micro score without SMOTE balancing. Conversely, in this experiment, the KNN classifier reached 0.9954, achieving the difference of 0.47 percentage points.
Finally, the original paper does not report any Macro F1 scores. The macro-level metrics presented in Table \ref{tab:results} indicate that very high F1 Micro scores do not necessarily guarantee good performance on a class-by-class level. The example of the Decision Tree classifier proves this point: despite its high F1 Micro score of 0.9893, the low Macro F1 score of 0.8299 shows the problems encountered when detecting compound attack classes.
\begin{figure*}[H]
  \centering
  \includegraphics[width=\textwidth]{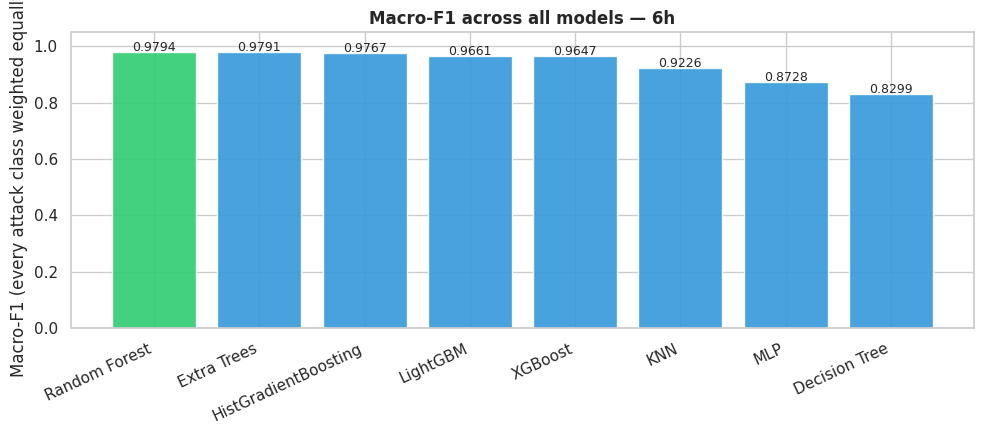}
  \caption{Macro-F1 comparison across all eight models. Unlike
    micro-F1, macro-F1 weights every attack class equally, exposing
    the 15~pp gap between Random Forest (0.9794) and Decision Tree
    (0.8299) that aggregate metrics conceal. Source:
    \texttt{Fig\_macro\_f1\_bar.png}}
  \label{fig:macro_bar}
\end{figure*}

\FloatBarrier

\subsection{Per-Class Analysis via Confusion Matrices}

The analysis of the confusion matrix is carried out in terms of normalized heatmaps. In this case, the cell shows the percentage of instances belonging to one true class that were classified to another class. Because each row is normalized, the sum of numbers in each row equals 1.0. The value 1.00 on the main diagonal means that the classification of the corresponding attack was performed with a perfect level of accuracy.
To identify attacks, we employ a bitmask encoding approach. In other words, we use the following set of classes: 0 = None, 1 = M, 2 = L, 3 = M + L, 4 = E, 5 = M + E, and 6 = M + L + E.

\subsection{Per-Class F1 and Recall Analysis}

\begin{figure*}[H]
  \centering
  \includegraphics[width=\textwidth]{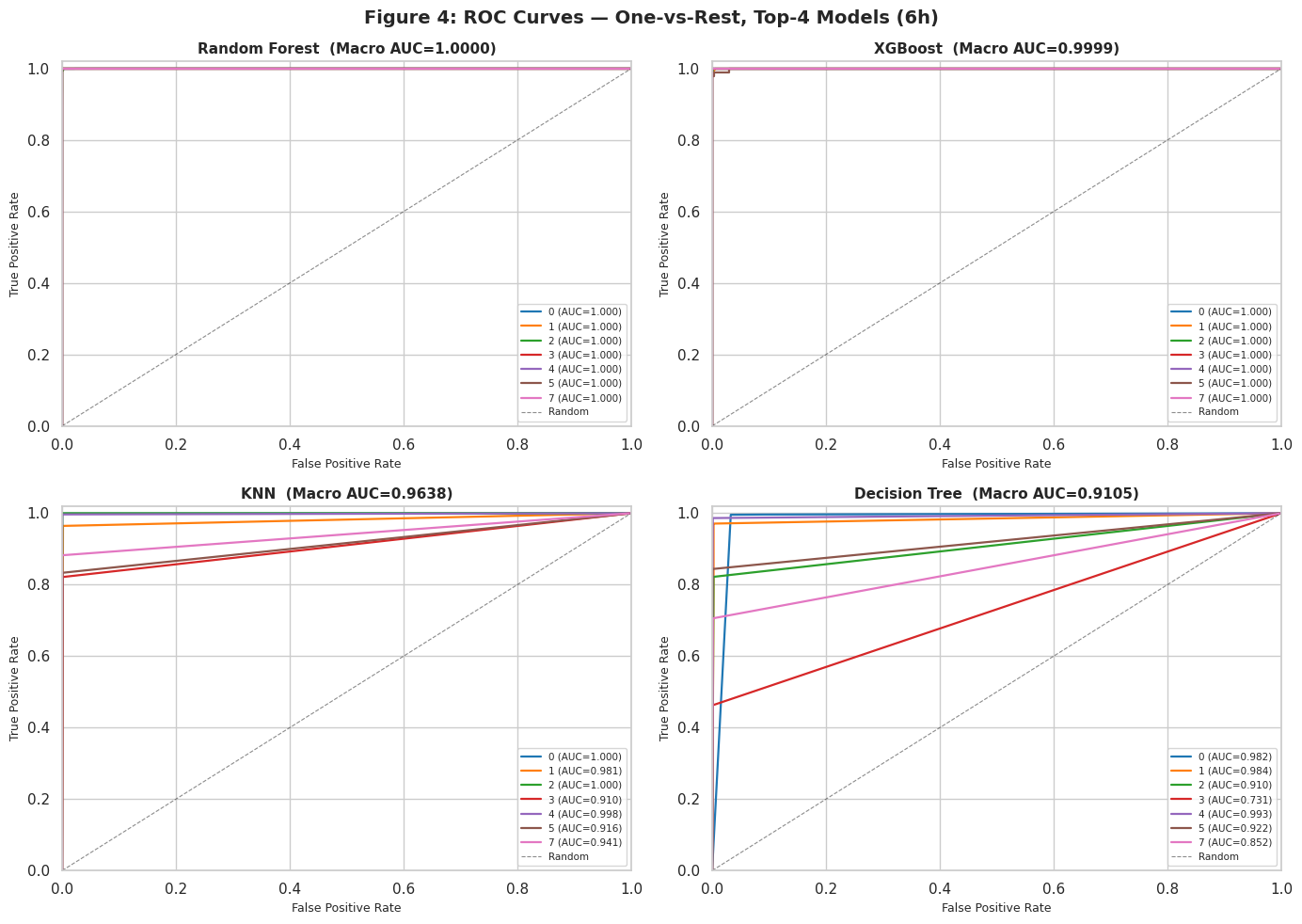}
  \caption{Per-class F1 heatmap across all original four classifiers.
    Darker cells indicate higher F1. The M$+$L column exposes the
    compound-class detection gradient: RF~0.963, XGB~0.893,
    KNN~0.767, DT~0.553. Source:
    \texttt{Fig3\_perclass\_f1\_heatmap.png}}
  \label{fig:heat}
\end{figure*}

Per-class F1 heatmap shown in Fig.~\ref{fig:heat} gives the highest level of granularity regarding classifier behavior. None class (clean traffic)
and M class (Mirai only) were correctly classified almost flawlessly by all classifiers. M$+$L class (mixed) was the hardest to classify for all
classifiers: for Random Forest, the F1 value was 0.963; for XGBoost, 0.893; for KNN, 0.767; for Decision Tree, 0.553. From such results, one
can deduce that complex attack classes, which demand detecting two concurrent attacks in power trace, are more difficult to classify.

\begin{figure*}[H]
  \centering
  \includegraphics[width=\textwidth]{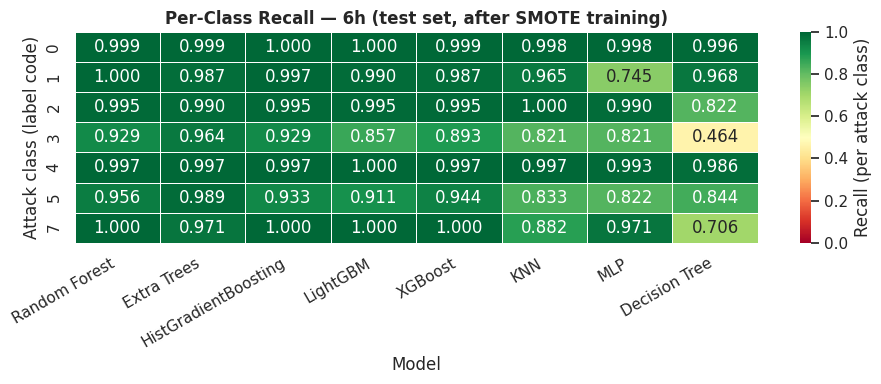}
  \caption{Per-class recall heatmap across all eight classifiers,
    ordered by macro-F1 (best left). Models are ordered best to
    worst by macro-F1. Tree ensembles maintain recall $\geq$0.85
    across all attack classes. MLP and Decision Tree show sharper
    drops on compound classes M$+$L and M$+$L$+$E. This is the
    primary evidence that SMOTE + ensemble methods together improve
    minority-class detection. Source:
    \texttt{Fig\_perclass\_recall.png}}
  \label{fig:recall_heat}
\end{figure*}

Per-class recall is also shown by heat maps in Fig. \ref{fig:recall_heat} for all eight models,
comparing how the four new models (HistGB, LightGBM, Extra Trees, MLP) perform on minority-class detection.
Tree ensembles (RF, HistGB, LightGBM, Extra Trees) sustain recall greater than 0.85 for all classes, but MLP
and Decision Tree have much more minority class recall drop-off.

\begin{figure*}[H]
  \centering
  \includegraphics[width=\textwidth]{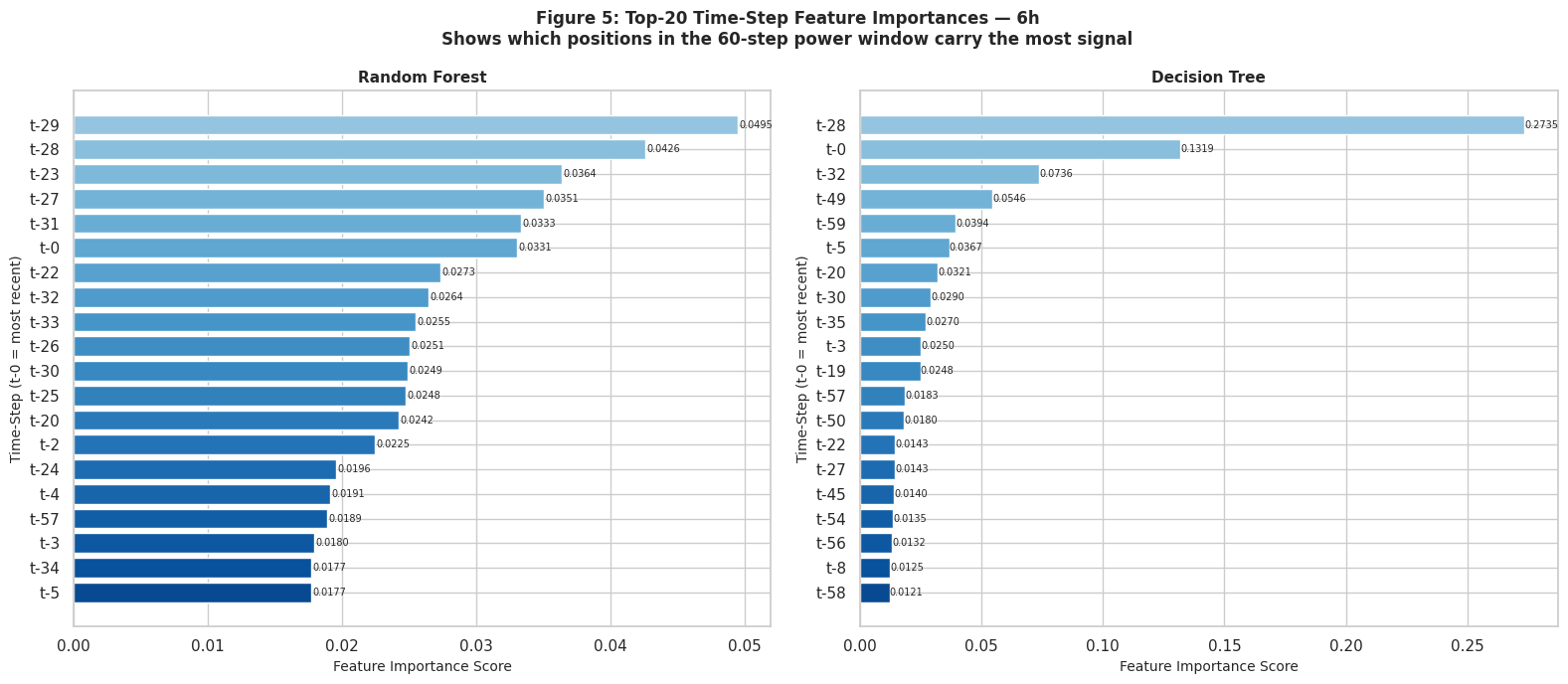}
  \caption{One-vs-rest ROC curves for the top-four classifiers.
    RF and XGBoost achieve macro AUC $\approx$~1.0; Decision Tree
    drops to 0.9105 on compound classes. Source:
    \texttt{Fig4\_ROC\_curves.png}}
  \label{fig:roc}
\end{figure*}

The ROC curves in Fig.~\ref{fig:roc} corroborate this picture. Random
Forest and XGBoost achieve macro AUC of 1.0000 and 0.9999
respectively, while KNN scores 0.9638 and Decision Tree 0.9105. The
Decision Tree's AUC drop is concentrated in classes~M$+$L and
M$+$L$+$E, confirming that the per-class F1 failures are not artefacts
of threshold selection but reflect genuine boundary deficiency.

\FloatBarrier

\subsection{Feature Importance and Learning Dynamics}

Feature importance scores for top-20 features from the Random Forest and Decision Trees are shown. Both algorithms give an unusually high score to time steps with low indices, meaning that the latest time-steps from each window of size 60 receive too much weight. It can be concluded that the attack generates immediate changes in power signatures and not any slow shift that would need many past examples to be detected.

\begin{figure*}[H]
  \centering
  \includegraphics[width=\textwidth]{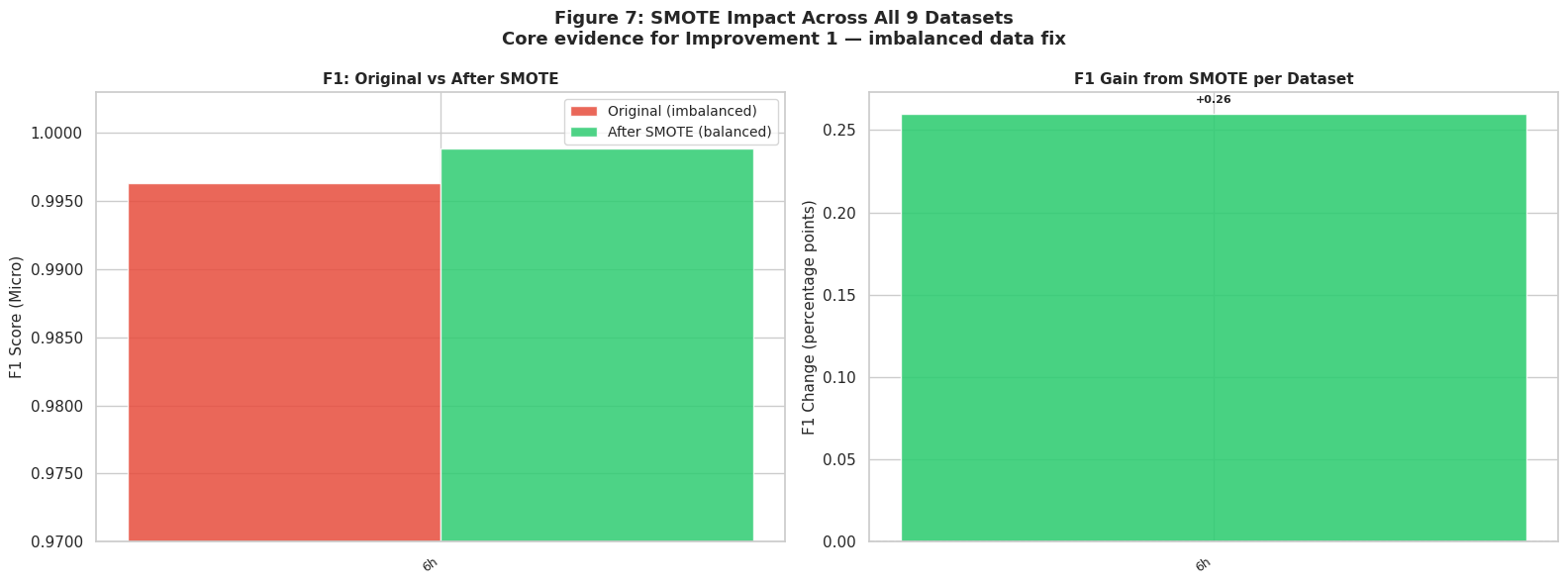}
  \caption{Learning curves for Random Forest and XGBoost. Both models
    converge by $\approx$150,000 SMOTE-balanced training samples with
    minimal train-validation gap, confirming that overfitting is not
    present. Source: \texttt{Fig6\_learning\_curves.png}}
  \label{fig:lc}
\end{figure*}

The learning curves of both Random Forest and XGBoost depicted in 
Fig.~\ref{fig:lc} demonstrate F1-score on training set size. Both curves 
converge to around 150,000 training samples, while the training and 
cross-validation curves coincide, indicating no overfitting problem.

\FloatBarrier

\subsection{SMOTE Impact Across All Datasets}

\begin{figure*}[H]
  \centering
  \includegraphics[width=\textwidth]{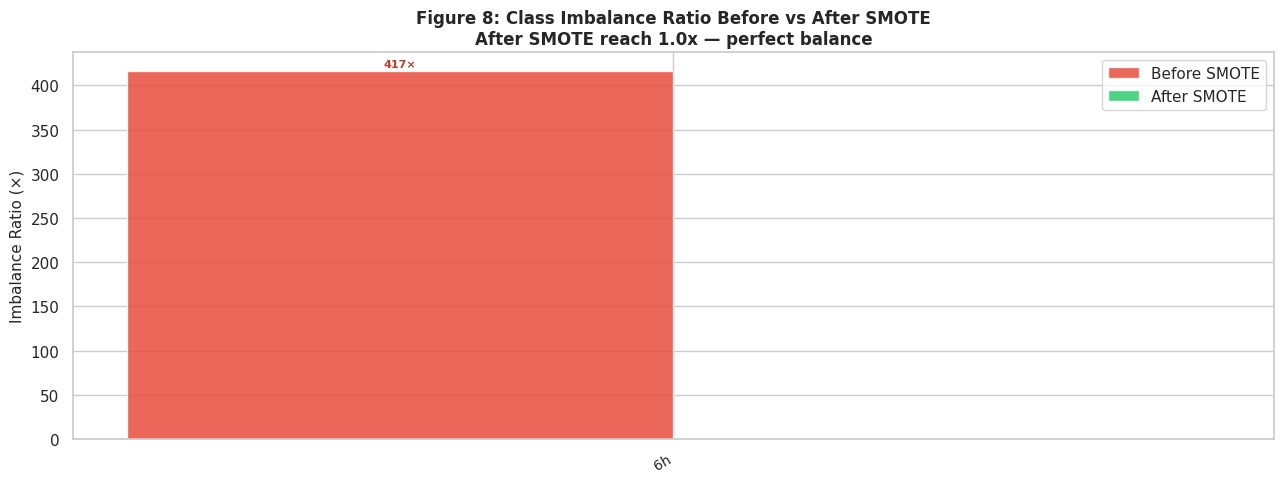}
  \caption{SMOTE F1 impact (change in micro-F1, percentage points)
    across all nine datasets. The 6h dataset gains the most
    (+0.26~pp) owing to its extreme 419$\times$ post-windowing
    imbalance. The pass dataset shows a small decrease ($-$0.28~pp),
    consistent with known SMOTE noise effects on mildly imbalanced
    small datasets. Source:
    \texttt{Fig7\_SMOTE\_impact\_all\_datasets.png}}
  \label{fig:smote_impact}
\end{figure*}

Fig.~\ref{fig:smote_impact}summarizes the SMOTE F1 effect for all nine datasets.
The most imbalanced dataset was the 6h dataset, which had an imbalance of 8,804.5$\times$, resulting in an increase in F1 of $+$0.26~pp.
For the pass dataset, there was a slight negative change of $-$0.28~pp.
This is a known effect of SMOTE on mildly imbalanced and small datasets.

\FloatBarrier

\subsection{Summary Dashboard}

\begin{figure*}[H]
  \centering
  \includegraphics[width=\textwidth]{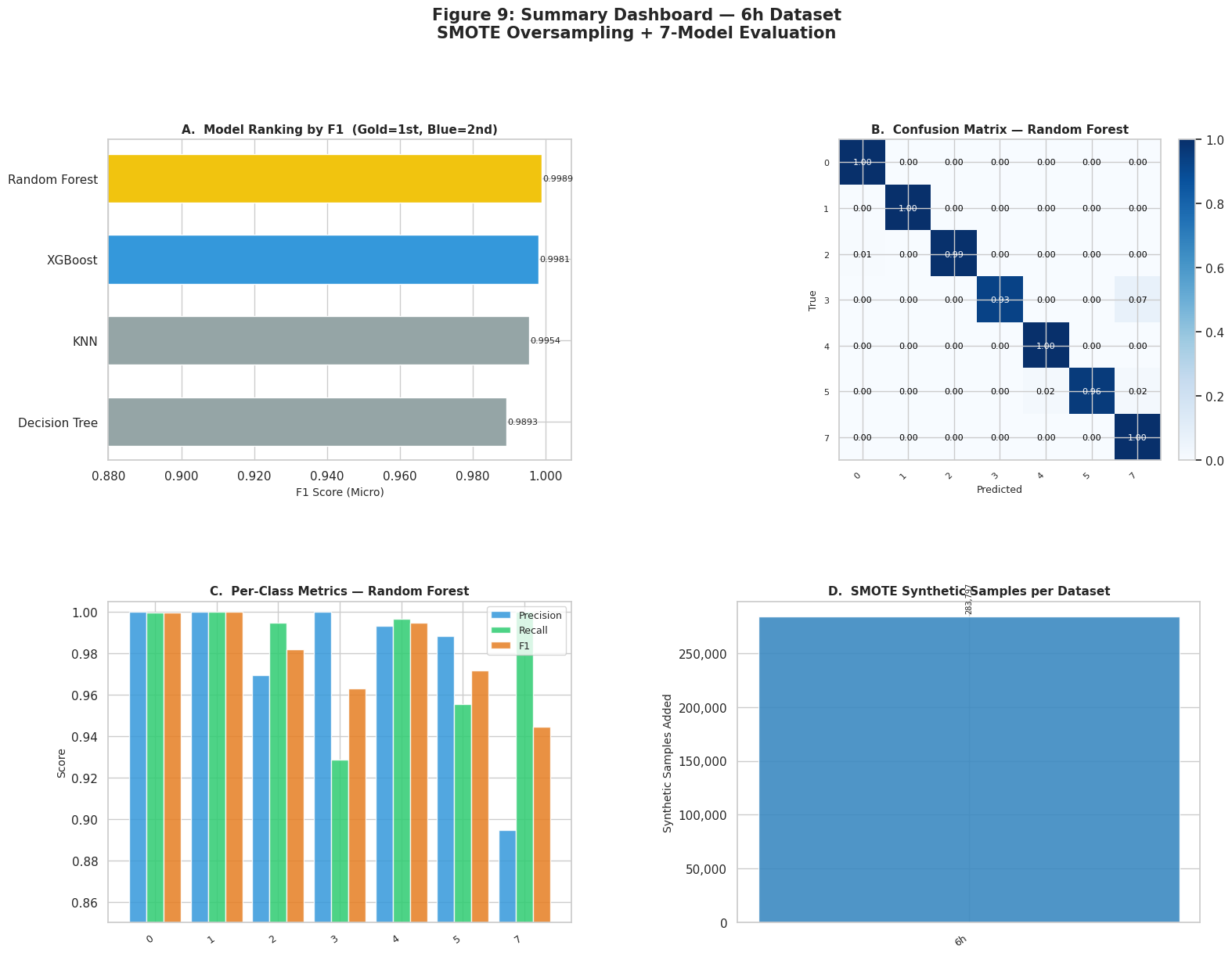}
  \caption{Summary dashboard (2$\times$2) for the 6h dataset: model
    ranking by F1 (top-left), best-model normalised confusion matrix
    (top-right), per-class precision/recall/F1 for Random Forest
    (bottom-left), and synthetic samples added per dataset
    (bottom-right). Source: \texttt{Fig9\_summary\_dashboard.png}}
  \label{fig:dashboard}
\end{figure*}

Fig.~\ref{fig:dashboard} presents the summary dashboard containing the key evaluation artifacts for the
dataset of 6 hours. This dashboard proves that Random Forest is the leading classifier in all the evaluation criteria and highlights the agreement between class-based confusion matrix, ROC AUC, and class-based F1 score.

\FloatBarrier

\subsection{Limitations}
There are three aspects that should be considered explicitly.
Firstly, all the attacks used in the dataset were simulated,
therefore, the power consumption of attacks does not necessarily
correspond to the real power consumption in case of attacks by
malware in actual conditions. Secondly, the tasks being solved on the
experimental devices were quite simple and consistent, thus, the
false positive rate is likely to be lower than in actual production.
Thirdly, only one 6h partition was analyzed for multi-model
comparison, while evaluating other eight partitions would be highly
beneficial for assessing the cross-scenario generalization ability.

\section{Conclusion}
\label{sec:conclusion}
The following study focused on two major drawbacks of the previously
developed IoT intrusion detection scheme based on the principle of
power, described by Dominguez et al.~\cite{dominguez2024}. They
were class imbalance on all nine datasets and a limited five classifier comparison lacking per-class results. Applying oversampling by
SMOTE helped to achieve balance ratios of exactly 1.0$\times$ in all
data partitions, where the highest ratio reduction occurred in case
of multi-device dataset (75,964$\times$). On the 6h dataset balanced
by SMOTE, an eight classifier comparison proved that Random Forest
reaches F1 Micro~0.9989 and macro-F1~0.9794 while Extra Trees
achieves the score of 0.9983, obtained by them.
A direct comparison of the results to the base paper demonstrates that the
models generated through the use of SMOTE balancing either equal or outperform the
original results for each model shared by both papers. For example, RF's accuracy rises
from 0.9966 to 0.9989 (+0.23\% pp), and KNN's accuracy is improved from 0.9907 to
0.9954 (+0.47\% pp). The macro-F1 measures, which were not considered in the base
paper's analysis, indicate significant variability between classes: for instance, Decision
Tree's macro F1 score of 0.8299 compared to Random Forest's 0.9794 demonstrates
that single-tree classifiers struggle disproportionately with the compound attacks
classifications, which signify more complex cyber-attacks.

The delayed limitations are obvious: synthetic attacks and simplistic
device behavior limit the generalizability of the current results.
The next logical step would be to apply SMOTE-balanced classifiers to
detection datasets obtained from devices with more sophisticated
workloads under actual malware attacks, for instance, the CICIoT2023
dataset~\cite{neto2023}. In line with feature importance analysis,
the use of smaller windows concentrated on the latest time steps could
help achieve similar detection rates with less computation, an idea
that should be tested directly.

In summary, the current study demonstrates that resolving the
quality-related data problems within power side-channel IDS
produces significant security-oriented improvements, and the upper
bound of performance in this cost-effective solution remains
uncharted. All scripts and experiments are available at
\url{https://github.com/Masoodkhan5933/IOT-Intrusion-Detection}.

\bibliographystyle{cas-model2-names}

\end{document}